\documentclass[a4paper]{PoS}
\usepackage{amsmath}
\usepackage{amssymb}
\usepackage{graphicx}
\usepackage{fontenc}
\usepackage{times}
\usepackage{mathptmx}
\newcommand{\apj}{{\it ApJ}} 
\newcommand{\apjl}{{\it ApJL}}
\newcommand{\mnras}{{\it MNRAS}}
\newcommand{\aap}{{\it A$\&$A}}
\newcommand{\ssr}{{Space Sci. Rev.,}}

\title{Similarity of $\gamma$-ray spectrum in middle aged supernova remnants (SNRs) interacting with molecular clouds (MC): what can we learn?}

\ShortTitle{$\gamma$-ray emission from old SNRs}

\author{\speaker{Xiaping Tang}\\
Max Planck Institute for Astrophysics, Karl-Schwarzschild-Str. 1, D-85741 Garching, Germany\\
        E-mail: \email{xt5uv@MPA-Garching.MPG.DE, tangxiaping@gmail.com}}

\abstract{In this work, we compare the $\gamma$-ray spectra available in literature from 11 middle aged supernova remnants (SNRs) interacting with molecular clouds (MCs). It is found that 5 remnants prefer a smoothly broken power law proton spectrum with similar power law index but different break energy. The rest of the SNRs need updated data to test whether a spectral break is preferred in the proton spectrum. Then we compare the $\gamma$-ray spectra from all 11 SNRs with the prediction from widely accepted escaping scenario and direct interaction scenario. We show that current $\gamma$-ray data is inconsistent with the escaping model statistically, as it predicts a diversity of $\gamma$-ray spectra which is not detected in the observation. We also find that ambient CRs can be very important for the $\gamma$-ray emission in the MCs external to W28 and W44, which requires further investigation. In the direct interaction scenario, we focus on re-acceleration of pre-existing ambient CRs. The model can produce the overall profile of $\gamma$-ray data with different acceleration time, but it suggests a transition of seed particles in the evolution of SNR. Whether such transition indeed exists has to be tested by future observation. In the end, we propose that radiative SNR without MC interaction can also produce a significant amount of $\gamma$-ray emission. One good candidate is S147. With accumulated Fermi data and CTA in future we expect to detect more remnants like S147. }

\FullConference{7th Fermi Symposium 2017\\
		15-20 October 2017\\
		Garmisch-Partenkirchen, Germany}

\begin{document}
\section{Introduction}
In the past few years, both space-based GeV observatories ({\it Fermi} and {\it AGILE}) and ground-based TeV observatories ({\it H.E.S.S, MAGIC} and {\it VERITAS}) detect $\gamma$-ray emission from several middle aged supernova remnants (SNRs) which are interacting with molecular clouds (hereafter SNR/MC) \cite{Acero16}. Multi-wavelength observations further demonstrate that the $\gamma$-ray emission region is spatially correlated with the MC interaction region \cite{Slane15}. The detected $\gamma$-ray emission is produced either by energetic electrons with Bremsstrahlung and Inverse Compton (IC) emission mechanism or accelerated protons with $\pi^0$-decay emission mechanism through proton-proton collision. 
Since dense MCs are ideal targets for proton-proton collision, the MC association established in observation supports a hadronic origin of the $\gamma$-ray emission. The characteristic $\pi^0$-decay signature around $67.5$MeV is considered to be the unique feature in spectra to distinguish hadronic emission from leptonic emission. Recently the identification of $\pi^0$-decay signature are proposed in SNRs like W44 and IC443 \cite{Ackermann13}, which is believed to be the first direct evidence for CR proton acceleration in SNRs. 

Despite above exciting progress in observation, our theoretical understanding about CR acceleration and emission in old SNRs are still very limited. It is partly because the evolution of old SNRs is more complicated and is strongly affected by the surrounding interstellar medium. 
Two scenarios are proposed to explain the observed $\gamma$-ray emission with hadronic origin and the association with MC interaction. One is the direct interaction scenario \cite{Uchiyama10,TC14}, in which the remnant is directly colliding with the MCs. The resulting interaction creates a cooling shock region with enhanced density and magnetic fields, where the CR protons and electrons are able to produce strong $\pi^0$-decay emission in $\gamma$-ray and synchrotron emission in radio respectively \cite{Uchiyama10}.
The other one is the escaping scenario \cite{Gabici09,LC10,Ohira11}, in which the MCs passively interact with the escaping CR particles from a adjacent SNR. Due to the high density and magnetic fields in the MCs, CR protons and electrons running into the clouds can illuminate them in $\gamma$-ray with $\pi^0$-decay emission and in radio with synchrotron emission respectively. 
In the following discussion, the non-thermal particles in the vicinity of a SNR are referred to as CR particles while the pre-existing CR background are instead referred to as ambient CRs.


\section{Comparison of $\gamma$-ray spectrum}
The growing number of SNR/MC detected in $\gamma$-ray enable us to investigate their physical properties in general and put better constrain on the theoretical models. Previous studies about $\gamma$-ray emission from SNR/MC focus on the individual source, in this work we compare the $\gamma$-ray spectra available in literature from 11 SNR/MC to obtain deeper insight into the physical origin of the emission. As a first attempt, in this study we focus on the shape of $\gamma$-ray spectra without providing detailed modeling for each individual SNR/MC. 

\begin{figure}
\includegraphics[width=1.02\textwidth]{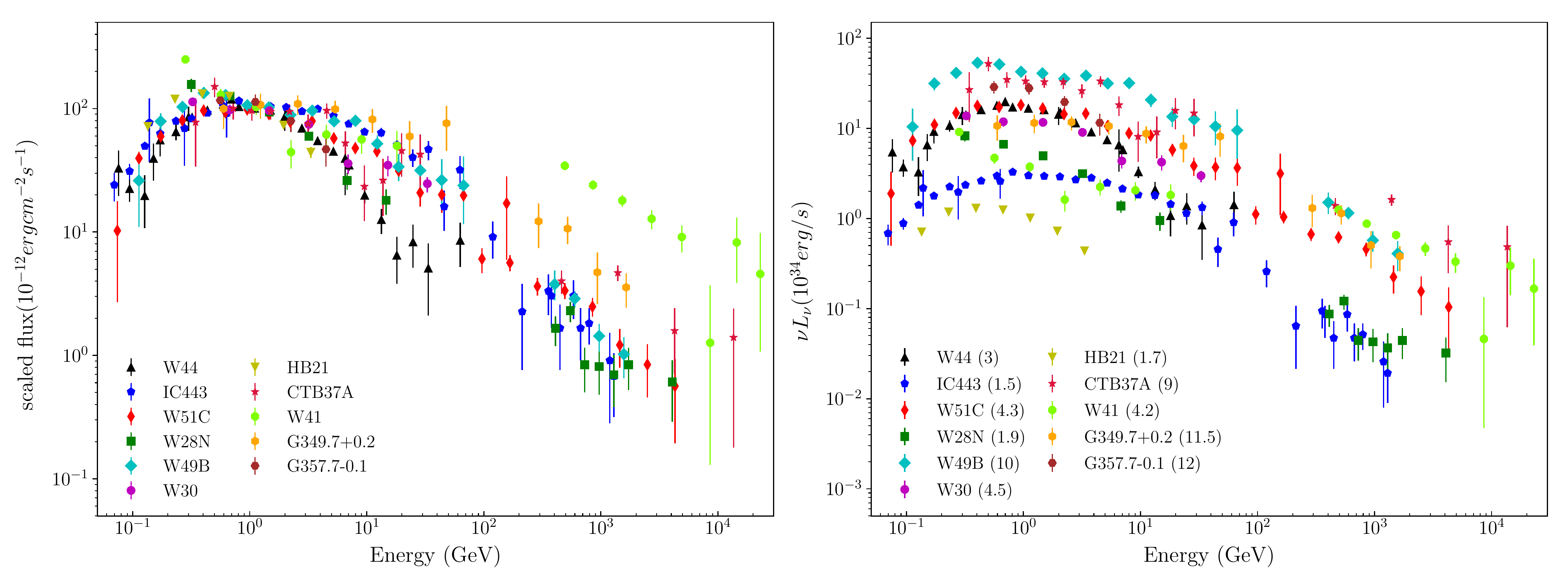}
\caption{Left panel: scaled $\gamma$-ray flux from 11 SNR/MC. Right panel: the corresponding $\gamma$-ray luminosity with distance in kpc indicated in the brackets. The reference for all the $\gamma$-ray data can be found in \cite{Tang17}. The error bars  are for statistical error only.}
\label{fig:SNR/MC}
\end{figure}

In the left panel of Fig. \ref{fig:SNR/MC}, we present the scaled $\gamma$-ray flux from 11 SNR/MC, which is normalized to have the same value around 1GeV. In the right panel, we plot the corresponding $\gamma$-ray luminosity with distance in kpc provided in the brackets. It is found that the spectra of many SNR/MC show a rising feature below  $\sim 1$GeV, which is consistent with $\pi^0$-decay emission from proton-proton collision. Above $\sim 1$GeV, most of the spectra follow a power law like profile with no clear sign for a high energy cutoff. The GeV data from space telescopes are smoothly connected with the TeV data from ground based observatories except W30. There is strong excess of TeV emission in W30, which is likely originated from a pulsar wind nebula \cite{Ajello12}. Hence, the TeV data of W30 is not illustrated in Fig. \ref{fig:SNR/MC}. The $\gamma$-ray luminosity from all 11 SNR/MC varies from $10^{34} \rm erg/s$ to $10^{36}\rm erg/s$. 
 
According to Fig. \ref{fig:SNR/MC}, several SNR/MC have similar spectral shape from sub-GeV all the way to TeV energies. To explore the similarity in the $\gamma$-ray spectra, we assume the $\gamma$-ray emission is dominated by $\pi^0$-decay and then fit the spectra with both a pow law and a smoothly broken power law (BPL) proton spectrum. The BPL proton spectrum as a function of momentum is assumed to be 
\begin{equation}
\frac{dN_p}{dp}\propto p^{-\alpha_1}\left[1+\left(\frac{p}{p_{br}} \right)^{(\alpha_2-\alpha_1)/w} \right]^{-w},
\end{equation}
where $p_{br}$ is the break momentum, $\alpha_1$ and $\alpha_2$ are power law index below and above the break respectively. $w$ determines the smoothness of the break and is fixed at 0.1 in the calculation. We found that 5 SNR/MC prefer a BPL proton spectrum and the results are presented in Table \ref{tab:fitting}. It is interesting to note that all 5 objects have similar $\alpha_1$ and $\alpha_2$ except W44, which has a much larger $\alpha_2$. The spectrum of W44 is much steeper than the rest of SNR/MC with no detection of TeV emission so far.  A possible explanation for the steep spectrum in W44 can be found in \cite{Uchiyama10,Tang17}.  The break momentum $p_{br}$ varies from several tens of GeV to hundreds of GeV.

\begin{table}
\begin{tabular}{ccccc}
\hline\hline
object& $\alpha_1$&$\alpha_2$& $p_{br}$ (GeV/c) &Reference for data used in fitting\\
\hline
IC443 &$2.28\pm 0.02$ &$3.27\pm 0.1$&$178\substack{+39\\-32}$  &\cite{Ackermann13}\\
W44 &$2.29\pm 0.06$ &$3.74 \substack{+0.17\\-0.15} $ & $33\substack{+6\\-5} $&\cite{Ackermann13} \\
W51C &$2.39\pm 0.03$ &$2.91 \pm 0.06$ &$112\substack{+32\\-35}$ &\cite{JF16}\\
W49B &$2.31\pm 0.03$ &$3.0\substack{+0.09\\-0.06}$ &$135 \substack{+68\\-32}$&\cite{HESS16}\\
G349.7+0.2&$2.21\substack{+0.2\\-0.15}$ &$2.74\pm 0.14$ &$180 \substack{+130\\-80}$ &\cite{HESS15a} \\
\hline
\end{tabular}
\caption{$\pi^0$-decay fitting results for 5 SNR/MC with BPL proton spectrum, which is calculated with the Naima Python package \cite{Zabalza15}.}
\label{tab:fitting}
\end{table}

Among the 6 SNR/MC left, G357.7-0.1, HB 21 and W30 have only a few data points in the GeV band, which is not good enough to test whether the proton spectrum has a spectral break. The TeV data of CTB 37A has large error bars which is also not able to put good constrain on the shape of proton spectrum. In W28N and W41, the low energy emission around $1$GeV appears to be inconsistent with $\pi^0$ decay, which implies the contribution of Bremsstrahlung emission might be important for them. Detailed modeling with updated {\it Fermi} data is needed in future to infer the proton spectrum in W28N and W41.

\section{Escaping scenario}
\begin{figure}
\includegraphics[width=1.02\textwidth]{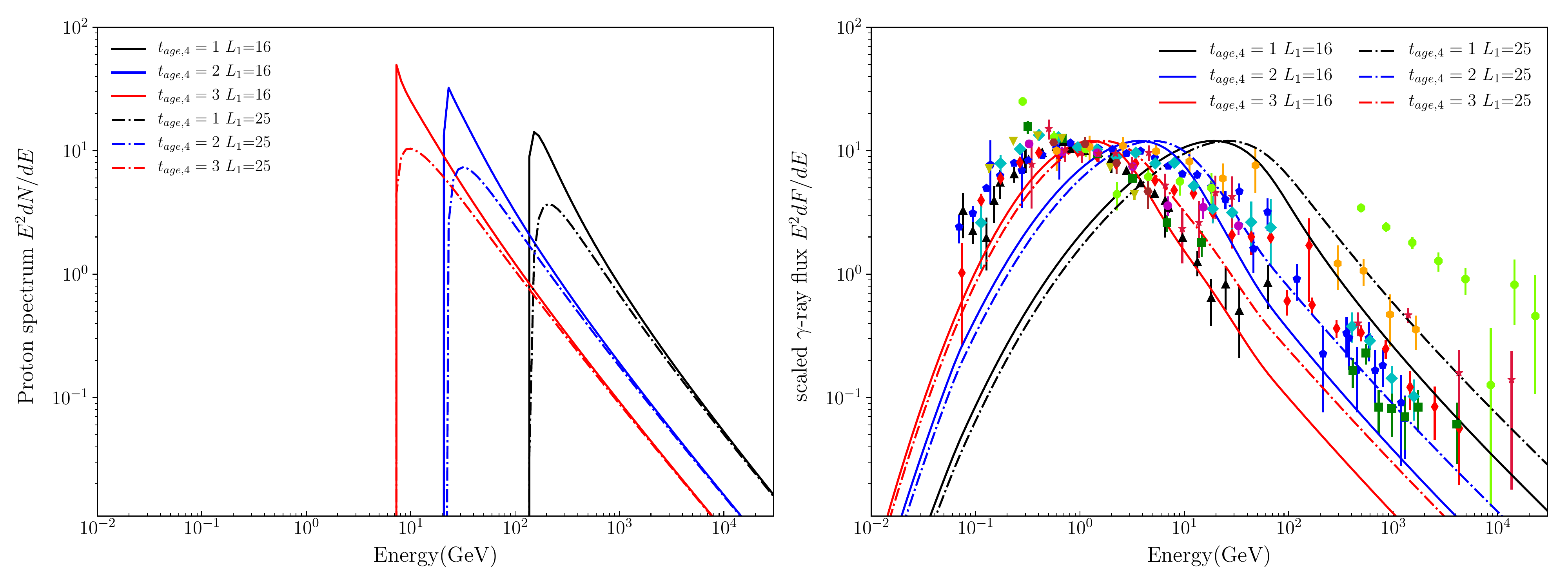}
\caption{Left panel: spectrum of escaping proton running into a nearby MC in arbitrary units. Right panel: the scaled $\pi^0$-decay emission from escaping model and the scaled $\gamma$-ray flux from observation.}
\label{fig:escape_model}
\end{figure}

In this section, we compare the escaping model prediction with observational data. In the left panel of Fig. \ref{fig:escape_model}, we plot the spectrum of escaping proton running into an adjacent MC with different cloud distance and remnant age. The escaping proton spectrum has a sharp low energy cutoff, which is due to the free escape boundary assumed in the model. In younger SNRs, the escaping proton spectrum is shifted to higher energy. Because only energetic particles are able to escape from the remnant in the early phase. In the right panel of Fig. \ref{fig:escape_model}, we compare the scaled $\pi^0$-decay emission from escaping model with the scaled $\gamma$-ray flux from observation. It is found that the escaping scenario can reproduce the overall profile of $\gamma$-ray data if only CR protons with energy as low as a few GeV are able to escape from the remnant and run into the nearby MCs (e.g., the red line). However, for observed SNR/MC with different remnant age and cloud distance, the model predicts a diversity of $\gamma$-ray spectra with different peak energy. The lack of such objects in observation (e.g., the black line) means that current data is inconsistent with the escaping scenario statistically. Besides, CR protons with energy around a few GeV are difficult to escape from the remnants  discussed here \cite{Tang17}. 

\begin{figure}
\includegraphics[width=1.02\textwidth]{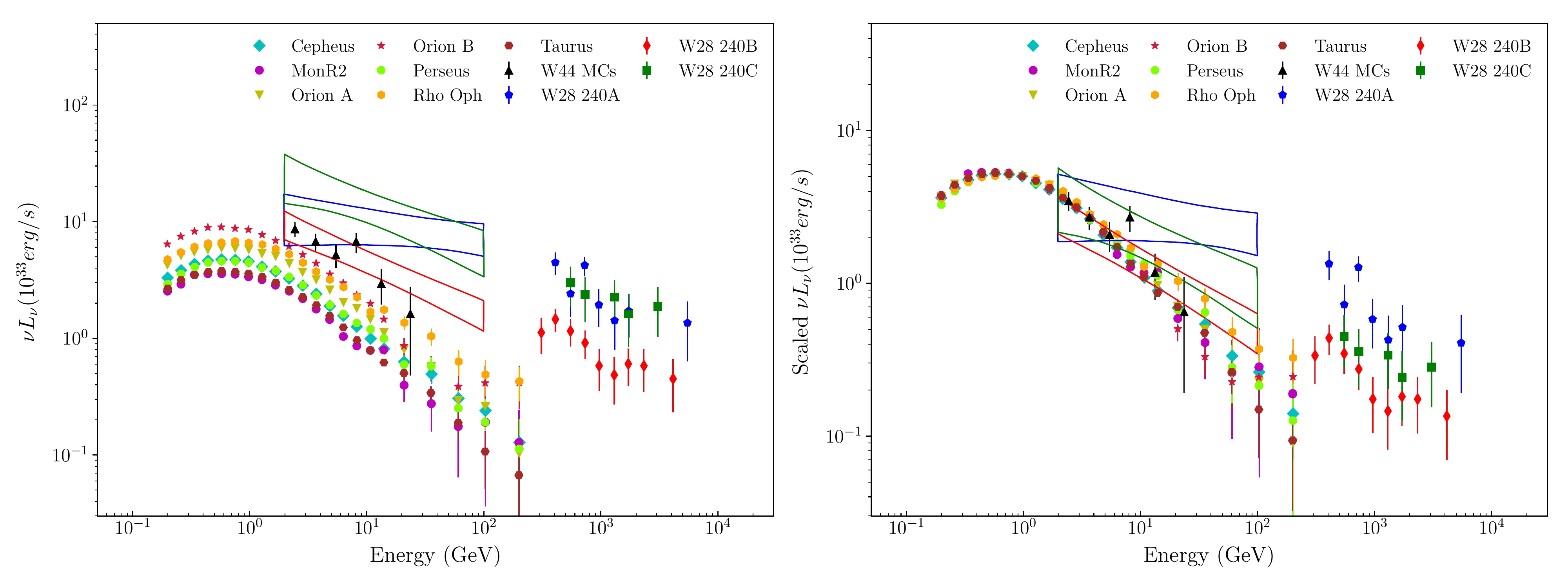}
\label{fig:GMC}
\caption{Left panel: $\gamma$-ray luminosity which is normalized for a cloud mass of $10^5 M_\odot$. Right panel: scaled $\gamma$-ray luminosity. The spectra for MCs in the Gould Belt is taken from \cite{Neronov17}.}
\end{figure}

The best example for escaping scenario is MCs adjacent to a SNR but is not spatially overlapped with the remnant. Two examples have been discussed extensively in the literature are HESS J1800-240 \cite{Aharonian08} external to W28 (hereafter W28 240) and the $\gamma$-ray bright clouds external to W44 (hereafter W44 MCs) \cite{Uchiyama12}. In the left panel of Fig. \ref{fig:GMC}, we present the $\gamma$-ray luminosity from W28 240, W44 MCs and several giant MCs in the Gould Belt. The luminosity is normalized for a cloud mass of $10^5 M_\odot$.  It is found that W28 240 and W44 MCs are indeed brighter than the isolated MCs in the Gould Belt. But the difference is not very significant considering the uncertainty in the cloud mass estimation. In the right panel of Fig. \ref{fig:GMC}, we scale the $\gamma$-ray luminosity to focus on their spectral shape. 
240A and 240C have harder spectrum in the very high energy comparing with MCs in the Gould Belt. However, 240A is spatially coincidented with two HII regions and 240C is spatially overlapped with SNR G5.71-0.08 which is likely interacting with MCs \cite{Hanabata14}. Whether the hardening is due to escaping CRs or spatially correlated source is still not clear. 
Based on above discussion, ambient CRs could be an important contributor to the $\gamma$-ray emission detected in W28 240 and W44 MCs, which need to be studied in a more careful way in the future.  

\section{Direct interaction scenario}
In \cite{Uchiyama10}, the authors studied the collision between a young (non-radiative) SNR and MCs. In our recent work, we discussed the situation when an old (radiative) SNR directly collides with MCs \cite{TC14}. In the young SNR case, only the MC interaction region can produce enhanced $\gamma$-ray and radio emission. In the old SNR case, not only the MC interaction region but also the radiative shell behind the remnant shock is able to produce enhanced $\gamma$-ray and radio emission. In \cite{TC14}, we propose that this new model can explain the discrepancy between radio and $\gamma$-ray morphology in IC 433. One interesting prediction from our model is that old (radiative) SNRs without MC interaction can also produce a significant amount of $\gamma$-ray emission \cite{Tang17}. One possible good example for this category is the nearby old remnant S147. 

Muti-wavelength observation show that the $\gamma$-ray, radio and H$\alpha$ emission in S147 are spatially correlated with each other \cite{Xiao08,Katsuta12}. The synchrotron radio and H$\alpha$ emission in SNRs trace the CR electrons and the cooling shell respectively. The $\gamma$-ray emission instead trace the CR protons, if we believe it is dominated by $\pi^0$-decay. The good correlation among $\gamma$-ray, radio and H$\alpha$ is consistent with the picture that relativistic protons and electrons are accumulated in the dense shell and then produce enhanced $\gamma$-ray and radio emission. Besides, there is no evidence for MC interaction in S147. The $\gamma$-ray luminosity of S147 is about $10^{33}\rm erg/s$ at $\sim 1$GeV with a distance of $1.3$kpc, which is smaller than the luminosity of SNR/MC presented in Fig. \ref{fig:SNR/MC} and is also consistent with a radiative SNR without MC interaction. 

\begin{figure}
\includegraphics[width=1.02\textwidth]{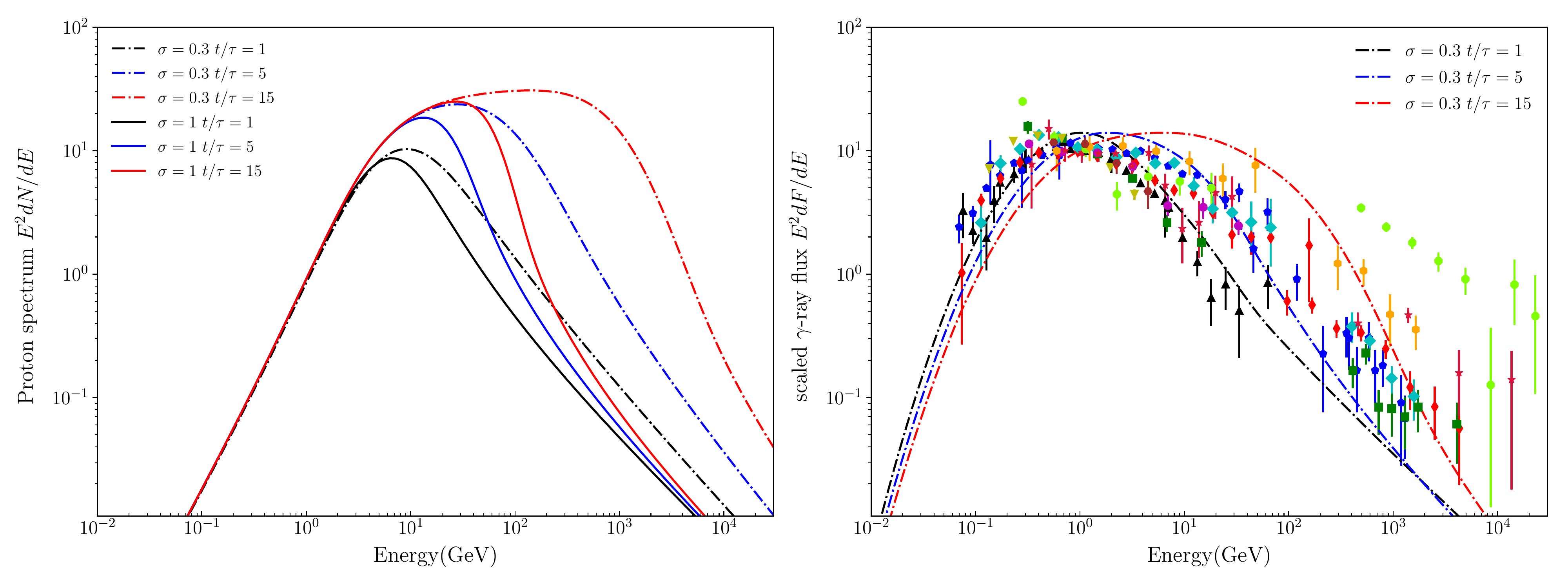}
\label{fig:interaction}
\caption{Left panel: proton spectrum after re-acceleration and adiabatic compression with different acceleration time and diffusion coefficients. Right panel: scaled $\pi^0$-decay emission from direct interaction model and the scaled $\gamma$-ray flux from observation.}
\end{figure}

In the direct interaction scenario, we focus on re-acceleration of ambient CRs, because the slow shock in the old SNRs are not able to accelerate thermal injected particles all the way up to TeV energy. If we consider only thermal injected particles, then it is difficult to explain the observed TeV emission. Another hint for re-acceleration of ambient CRs is from the observed $\gamma$-ray spectra. The BPL proton spectrum presented in Table \ref{tab:fitting} is close to the ambient CR proton spectrum derived in \cite{Neronov17} with $\alpha_1= 2.33\substack{+0.06\\-0.08}, \,\alpha_2=2.92\substack{+0.07\\-0.04}$  and $p_{br}=18.35\substack{+6.48\\-3.57}$. The main difference is in the break momentum which is possibly due to different acceleration time. In the left panel of Fig. \ref{fig:interaction}, we plot the proton spectrum after re-acceleration and adiabatic compression with different acceleration time and diffusion coefficients. The resulting proton spectrum is characterized by a momentum break which is determined by the acceleration time \cite{TC15}. Below the break, the accelerated proton spectrum is consistent with the steady state solution of diffusive shock acceleration \cite{Bell78}, while above the break the spectrum instead follows the shape of ambient CR. When the acceleration time increases, the break is shifted to higher energy as CR proton with higher energy are able to enter the steady state. In the right panel of Fig. \ref{fig:interaction}, we compare the scaled $\pi^0$-decay emission from models with the scaled $\gamma$-ray flux from observation. It is found that the model can reproduce the overall shape of $\gamma$-ray data with different acceleration time. The main issue about re-acceleration is that it implies a transition of seed particles in the SNR evolution, which is from thermal injected seed particles in young SNRs to ambient CRs in old SNRs, please see \cite{Tang17} for detailed discussion

\section{Discussion}
We compare the $\gamma$-ray spectra from 11 SNR/MC with prediction from theoretical models. We found that the $\gamma$-ray data is inconsistent with the escaping scenario statistically. The inconsistency is mainly due to the free escape boundary assumed in the model, which implies free escape boundary may not be a good recipe to study CR escaping. We also show that ambient CRs could be an important contributor to the $\gamma$-ray emission in the illuminated clouds external to W44 and W28. The direct interaction scenario involving re-acceleration 
of ambient CRs can reproduce the overall profile of $\gamma$-ray data. But the model implies a transition of seed particles in the evolution of SNRs, which requires further investigation. In the end, we propose that old (radiative) SNRs without MC interaction are also able to produce a significant amount of $\gamma$-ray emission. 

\section*{Acknowledgments}
X.T thank the organizing committee for organizing this stimulating Fermi symposium.

\end{document}